# Estimation of Timing Offsets and Phase Shifts Between Packet Replicas in MARSALA Random Access

Karine Zidane*, Jérôme Lacan*, Mathieu Gineste§, Caroline Bes¶, Arnaud Deramecourt¶ and Mathieu Dervin§

*Univesity of Toulouse, ISAE/DISC & TéSA

Email: {karine.zidane, jerome.lacan}@isae.fr

§Thales Alenia Space, Toulouse

Email: {mathieu.gineste, mathieu.dervin}@thalesaleniaspace.com

¶CNES, Toulouse

Email: {caroline.bes, arnaud.deramecourt}@cnes.fr

*Abstract*—Multi-replicA decoding using corRelation baSed LocALisAtion (MARSALA) is a recent random access technique designed for satellite return links. It follows the multiple transmission and interference cancellation scheme of Contention Resolution Diversity Slotted Aloha (CRDSA). In addition, at the receiver side, MARSALA uses autocorrelation to localise replicas of a same packet so as to coherently combine them. Previous work has shown good performance of MARSALA with an assumption of ideal channel state information and perfectly coherent combining of the different replicas of a given packet. However, in a real system, synchronisation errors such as timing offsets and phase shifts between the replicas on separate timeslots will result in less constructive combining of the received signals. This paper describes a method to estimate and compensate the timing and phase differences between the replicas, prior to their combination. Then, the impact of signal misalignment in terms of residual timing offsets and phase shifts, is modeled and evaluated analytically. Finally, the performance of MARSALA in realistic channel conditions is assessed through simulations, and compared to CRDSA in various scenarios.

## I. INTRODUCTION

On satellite return links, Demand Assignment Multiple Access (DAMA) methods allow the user terminals to access the satellite ressources using a set of carrier frequencies divided into timeslots. In a system with bursty internet traffic and relatively short packets, DAMA methods would induce ressource request/allocation delays and might not be even practical for very large populations of terminals. In this context, the use of Random Access (RA) methods possibly in association with DAMA for data transmission, would be of interest.

Legacy synchronous RA protocols (i.e. Slotted Aloha (SA) [1], Diversity Slotted Aloha (DSA) [2]) offer rather poor performance due to the fact that packet collisions are often destructive. New protocols have arised to improve the performance of synchronous RA. For this, these recent protocols apply the principle of Successive Interference Cancellation (SIC), besides redundancy transmission.

Among these techniques, Contention Resolution Diversity Slotted Aloha (CRDSA) [3] has been included as an option in the recent standard DVB-RCS2 [4], [5]. In CRDSA, each user sends multiple replicas of the same packet on different timeslots of the frame. Each copy contains pointers towards the next replicas. The receiver performs interference cancellation each time a packet is decoded successfully. In the first version of CRDSA, each user transmits only two copies per packet, and all the terminals are supposed to transmit at equal power. The performance of CRDSA has been enhanced with CRDSA++ [6] where more than two replicas per packet can be transmitted by each user. CRDSA++ also exploits the received packets power unbalance. Irregular Repetition Slotted Aloha (IRSA) has been proposed in [7] as a generalized version of CRDSA. In IRSA, the terminals can send different numbers of packet replicas, in order to enhance the diversity on a given frame.

Another RA technique proposed in [8] is Multi-Slot Coded Aloha (MuSCA). In MuSCA, instead of sending replicas of the same packet, the transmitter encodes the packet with a robust forward error correction (FEC) code of rate $R$, then divides the code word into multiple fragments and adds seperately coded signalling fields to localise each fragment on the frame. The receiver first decodes the signalling information in order to localise the fragments associated to one packet, then combines the fragments together to reconstruct the codeword. MuSCA significantly improves the system performance, at the cost of increased signalling overhead and incompatibility with the DVB-RCS2 standard.

In order to further enhance the RA performance and to be compatible with the DVB-RCS2 standard, another RA method named Multi-ReplicA Decoding using corRelation baSed LocAlisAtion (MARSALA) has been introduced in [9]. MARSALA proposes a new decoding technique for CRDSA based on replicas localisation using autocorrelation. In particular, MARSALA takes advantage of correlation procedures between the signals received on different timeslots, in order to locate the replicas of one packet even when all of them are undergoing a collision and their pointers are not decodable. Furthermore, the coherent combination of the identified packet replicas allows to enhance the Signal to Noise plus Interference Ratio (SNIR), in order to possibly enable decoding and trigger additional SIC steps. The transmitter side in MARSALA is the same as in CRDSA, the only modifications are at the receiver side. MARSALA does not require any implementation modifications to the DVB-RCS2 standard.

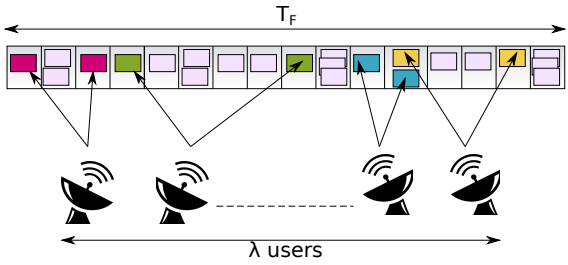

Fig. 1: System model with $\lambda$ terminals and $N_b = 2$ replicas per packet

In previous work [10], [11], we have analyzed the effect of residual channel estimation errors on the performance of RA methods that use the SIC principle. In MARSALA, an additional critical task is to estimate and compensate for the synchronisation differences between the replicas on separate timeslots, so as to allow their coherent combination. This step is very important to maximize the combination gain. There are two main contributions in this paper: a method to estimate and compensate the timing offsets between packet replicas, as well as the phase shifts caused by carrier frequency and phase variations. And the definition of an analytical model of the performance degradation caused by imperfect replicas combination due to the estimation errors. This model is being used to evaluate performance degradation of the scheme and compare MARSALA in realistic channel conditions, to CRDSA.

## II. SYSTEM HYPOTHESIS

We consider the Multi-Frequency Time Division Multiple Access (MF-TDMA) Return Link structure defined in DVB-RCS2 standard for GEO satellite systems. Our system model considers one RA channel over one frequency carrier with frames divided into $N_s$ timeslots each, and shared between $\lambda$ users. The users are the Return Channel over Satellite Terminals (RCSTs), and are supposed to be fixed terminals. Each user transmits $N_b$ copies of the same packet to a destination node (a satellite or a gateway) over different timeslots within the duration of one frame ($T_F$). We suppose that, to send other packets, the user must wait until the beginning of the next frame. We assume that all the received packets are equi-powered.

Each user experiences independant, random timing offsets, frequency offsets and phase shifts, distributed as follows: The timing offset is uniformly distributed in $[-T_s, T_s]$ and varies from one timeslot to another, with $T_s$ being the symbol period; the carrier frequency offset is supposed to remain constant for one user over the duration of one frame and is uniformly distributed in $[0, \frac{0.01}{T_s}]$; the phase shift is uniformly distributed in $[-\pi, \pi]$ and may vary from one timeslot to another.

We consider an Additive White Gaussian Noise (AWGN) channel model. The AWGN power spectral density denoted by $N_0$ is constant over $T_F$. The channel amplitude is supposed normalized to 1. Each packet is encoded with the DVB-RCS2 turbo-encoder for linear modulation, of rate $R$. The resulting code word is modulated with a modulation of order $M$. A preamble and a postamble are added at the beginning and at the end of each packet, and pilot symbols are distributed inside the packet for the purpose of channel estimation. The total packet length is equal to $L$ symbols. Before the transmission, the symbols corresponding to each packet enter a shaping filter with a square root raised cosine function.

At the receiver side, the frame is processed using CRDSA. When all the packets are in a non-decodable situation, the receiver applies MARSALA in order to attempt to decode additionnal packets and possibly trigger more SIC iterations. In MARSALA, the receiver identifies one timeslot as a reference timeslot denoted by $TS_{ref}$. The signal received on $TS_{ref}$ is expressed as follows

$$r_1(t) = y(t + \tau_1)e^{j(\phi_1 + 2\pi\Delta f_1 t)} + n_1(t) + \zeta_1(t) \quad (1)$$

with $y$ being the useful signal to demodulate and decode. $\tau_1$, $\phi_1$, and $\Delta f_1$ denote respectively, the timing offset, the phase shift, and the frequency offset relative to $y$ on $TS_{ref}$. $n_1$ is the AWGN term and $\zeta_1$ represents the total interferent signals on $TS_{ref}$. The signal $y$ can be detailed as

$$y(t) = \sum_{k=0}^{L-1} a_k h(t - kT_s) \quad (2)$$

with $a_k$ being the $k^{th}$ symbol corresponding to $y$, and $h$ denoting the shaping filter function. As for the remaining $(N_b - 1)$ replicas of $y$, their corresponding signal can be expressed as

$$r_i(t) = y(t - N_i T_s + \tau_i)e^{j(\phi_i + 2\pi\Delta f_1 t)} + n_i(t) + \zeta_i(t) \quad (3)$$

with $i$ being the integer index identifying each replica, $i \in [2 : N_b]$. $N_i$ is the number of symbols separating the useful packet on $TS_{ref}$ from its $i^{th}$ replica. $\tau_i$ and $\phi_i$ refer respectively to the timing offset and the phase shift relative to $y$ on the timeslot containing the $i^{th}$ replica. $n_i$ and $\zeta_i$ are respectively, the AWGN term and the total interferent signals on the timeslot containing the $i^{th}$ replica.

## III. ESTIMATION AND COMPENSATION OF THE TIMING OFFSET AND PHASE SHIFT BETWEEN REPLICAS

This section aims to describe a synchronisation scheme for the received packet replicas in MARSALA. The main objective is to be able to perform coherent replicas combination. In the following, we first recall the method for replicas localisation using a correlation based technique. Then, we describe the procedure to estimate and compensate the timing offsets and phase shifts between localised replicas.

### A. Replicas localisation using a correlation based technique

To localise the replicas of a given packet, the receiver computes the cross-correlation between $r_1$ and the signals on the rest of the frame. The correlation peaks indicate the locations of the replicas of a packet present on $TS_{ref}$. The signal $r_i$ described in Section II, can be expressed using $r_1$, $n_1$, $\zeta_1$, $n_i$ and $\zeta_i$ as shown below

$$r_i(t) = r_1(t - N_i T_s + \Delta\tau_{i,1})e^{j\Delta\phi_{i,1}} + n_{tot}(t) \quad (4)$$

where $N_i T_s - \Delta\tau_{i,1} = N_i T_s - (\tau_i - \tau_1)$ is the timing offset between the packet on $TS_{ref}$ and its $i^{th}$ replica. $\Delta\phi_{i,1} = \phi_i - \phi_1 + 2\pi\Delta f_1(N_i T_s + \Delta\tau_{i,1})$ is the phase shift between the two replicas. $n_{tot}$ is the total noise plus interference term expressed as follows.

$$n_{tot}(t) = [n_i(t) + \zeta_i(t)] - [n_1(t - N_i T_s + \Delta\tau_{i,1}) \\ + \zeta_1(t - N_i T_s + \Delta\tau_{i,1})]e^{j\Delta\phi_{i,1}} \quad (5)$$

Without loss of generality, we will only express the cross-correlation function between $r_1$ and $r_i$. Since all the signals are supposed to have finite power, the correlator output can be expressed as follows

$$R_{r_i,r_1}(\tau) = \lim_{T \to \infty} \frac{1}{T} \int_{-T}^{T} r_i(t) r_1^*(t-\tau) dt$$

$$= \lim_{T \to \infty} \frac{1}{T} \int_{-T}^{T} r_1(t - N_i T_s + \Delta\tau_{i,1}) r_1^*(t-\tau) e^{j\Delta\phi_{i,1}} dt$$

$$+ \lim_{T \to \infty} \frac{1}{T} \int_{-T}^{T} n_{tot}(t) r_1^*(t-\tau) dt$$

$$= R_{r_1}\left(\tau - (N_i T_s - \Delta\tau_{i,1})\right) e^{j\Delta\phi_{i,1}} + R_{n_{tot},r_1}(\tau) \quad (6)$$

with $R_{r_1}$ denoting the autocorrelation function of $r_1$ and $R_{n_{tot},r_1}$ being the cross-correlation function between $n_{tot}$ and $r_1$. The operator $(\cdot)^*$ denotes complex conjugate.

The absolute value of $R_{r_i,r_1}$ reaches its maximum at $\tau = \tau_{max_i} = N_i T_s - \Delta\tau_{i,1}$, which represents the timing offset between the first and the $i^{th}$ replica of the same packet. For $\tau = \tau_{max_i}$, $R_{r_i,r_1}(\tau_{max_i})$ is expressed as shown below, with $P_{r_1}$ being the power of the signal $r_1$.

$$R_{r_i,r_1}(\tau_{max_i}) = P_{r_1} e^{j\Delta\phi_{i,1}} + R_{n_{tot},r_1}(\tau_{max_i}) \quad (7)$$

### B. Replicas synchronisation procedure

In order to combine $r_1$ and $r_i$ coherently, proper timing and phase synchronisation between the replicas is required. Therefore, the receiver shall estimate the timing offset $\tau_{max_i}$ and the phase shift $\Delta\phi_{i,1}$ caused by carrier frequency and phase variations between $r_1$ and $r_i$. In a real system, the received signal is sampled with an oversampling factor $Q$. If $\tau_{max_i}$ is not an integer multiple of $\frac{T_s}{Q}$, the correlation peak would be detected with a timing error $err_i$. The timing error can be modeled with a random variable uniformly distributed in $\left[\frac{-T_s}{2Q}, \frac{T_s}{2Q}\right]$. Then, the estimate of the timing offset is expressed as $\widehat{\tau}_{max_i} = (N_i T_s - \Delta\tau_{i,1} + err_i)$. The value of the cross-correlation function $R_{r_i,r_1}$ at $\widehat{\tau}_{max_i}$ is

$$R_{r_i,r_1}(\widehat{\tau}_{max_i}) = R_{r_1}(err_i) e^{j\Delta\phi_{i,1}} + R_{n_{tot},r_1}(\widehat{\tau}_{max_i})$$

$$= \left[\frac{H(err_i)}{T_s} e^{j\Delta f_1 err_i}\right] e^{j\Delta\phi_{i,1}}$$

$$+ \lim_{T \to \infty} \frac{1}{T} \int_{-T}^{T} y(t + \tau_1) n_1^*(t - err_i) e^{j(\phi_1 + 2\pi\Delta f_1 t + \Delta\phi_{i,1})} dt$$

$$+ \lim_{T \to \infty} \frac{1}{T} \int_{-T}^{T} y(t + \tau_1) \zeta_1^*(t - err_i) e^{j(\phi_1 + 2\pi\Delta f_1 t + \Delta\phi_{i,1})} dt$$

$$+ R_{n_i,r_1}(\widehat{\tau}_{max_i}) + R_{\zeta_i,r_1}(\widehat{\tau}_{max_i}) \quad (8)$$

with $H$ being the raised cosine filter function.

Given that $\Delta f_1$ is supposed to be constant over the duration of one frame, then the phase shift between replicas $\Delta\phi_{i,1}$ is also constant. Therefore, $\widehat{\Delta\phi}_{i,1} = \text{angle}\left(R_{r_i,r_1}(\widehat{\tau}_{max_i})\right) = (\Delta\phi_{i,1} + \phi_{err_i})$, with $\phi_{err_i}$ being a phase error that depends on $\Delta f_1 err_i$, $H(err_i)$ and the cross-correlation functions between noise, interference and the useful signal. We can conclude from Eq. (8), that in order to minimize the error $\phi_{err_i}$, we shall choose a reference signal $r_1$ containing the lowest interference level. Since all the replicas of a same packet are equi-powered, then among the signals corresponding to replicas of a same packet, the signal having the lowest total received power on one timeslot shall have the lowest interference power of $\zeta_1$. In other words, in order to estimate $\Delta\phi_{i,1}$ after replicas localisation, the receiver shall choose $r_1$ as the received signal on one localised timeslot with the lowest power level.

## IV. DEFINITION OF AN ANALYTICAL MODEL FOR REPLICAS COMBINATION WITH SYNCHRONISATION ERRORS

After estimation and compensation of timing and phase according to the scheme described above, replicas combination is performed. However, residual timing offsets and phase shifts are added to the combined replicas. Therefore, we define an analytical model for the impact of imperfect replicas combination on the performance of MARSALA.

### A. Replicas Combination Scheme

The signal $r_1$ on $TS_{ref}$ is sampled at time instants $t = n\frac{T_s}{Q}$, with $n$ being an integer varying from $0$ to $(Q * L) - 1$. The resulting discrete signal is expressed as follows.

$$r_1(n) = y\left(n\frac{T_s}{Q} + \tau_1\right) e^{j\left(\phi_1 + 2\pi\Delta f_1 n \frac{T_s}{Q}\right)} + n_1\left(n\frac{T_s}{Q}\right) + \zeta_1\left(n\frac{T_s}{Q}\right) \quad (9)$$

The signal $r_i$ is sampled at instants $t = n\frac{T_s}{Q} + \widehat{\tau}_{max_i}$. Then, the phase shift is corrected with respect to the phase and frequency offsets of $r_1$, by multiplying the resulting samples with $e^{-j\widehat{\Delta\phi}_{i,1}}$. The corresponding discrete signal is expressed as shown below.

$$r_i(n) = y\left(n\frac{T_s}{Q} + \tau_1 + err_i\right) e^{j\left(\phi_1 + \phi_{err_i} + 2\pi\Delta f_1 n\frac{T_s}{Q}\right)}$$
$$+ \left[n_i\left(n\frac{T_s}{Q} + \widehat{\tau}_{max_i}\right) + \zeta_i\left(n\frac{T_s}{Q} + \widehat{\tau}_{max_i}\right)\right] e^{-j\widehat{\Delta\phi}_{i,1}} \quad (10)$$

Replicas combination consists in the summation of $r_1$ and $r_i$ for $i = [2, \ldots, N_b]$, then matched filtering and downsampling the resulting signal at instants $k'T_s$, with $k'$ being an integer varying from 0 to $L - 1$. In the following, we will denote by $z_1$ and $z_i$ the noise plus interference term in signals $r_1$ and $r_i$, respectively. The resulting discrete signal is expressed as follows

$$r_{sum}(k'T_s) = y_{sum}(k'T_s)e^{j(\phi_1 + 2\pi\Delta f_1 k'T_s)} + \tilde{z}_1(k'T_s)$$
$$+ \sum_{i=2}^{N_b} \tilde{z}_i(k'T_s + \widehat{\tau}_{max_i})e^{-j\widehat{\Delta\phi}_{i,1}} \quad (11)$$

where $\tilde{(\cdot)}$ denotes the signal at the output of the matched filter, and $y_{sum}$ is detailed as shown below.

$$y_{sum}(k'T_s) = \tilde{y}(k'T_s + \tau_1)$$
$$+ \sum_{i=2}^{N_b} \tilde{y}(k'T_s + \tau_1 + err_i) e^{j\phi_{err_i}} \quad (12)$$

Since $y_{sum}$ is the sum of signals sampled at imperfect sampling times, it can be divided into two terms: a desired signal denoted by $y_{sum,des}$ and an Inter-Symbol Interference (ISI) term denoted by $y_{sum,isi}$. For the ISI term, we consider only the first two side lobes on either side of the raised cosine filter, since the interferent symbol would be attenuated by more than 10 dB beyond these lobes. Both terms are detailed below

$$y_{sum,des}(k'T_s) = a_{k'}H(\tau_1) + \sum_{i=2}^{N_b} a_{k'}H(\tau_{err_i})e^{j\phi_{err_i}} \quad (13)$$

$$y_{sum,isi}(k'T_s) = \sum_{l=k'-3, l\neq k'}^{k'+3} a_l H((k'-l)T_s + \tau_1)$$
$$+ \sum_{i=2}^{N_b} \left[\sum_{l=k'-3, l\neq k'}^{k'+3} a_l H((k'-l)T_s + \tau_{err_i})\right] \quad (14)$$

with $a_{k'}$ and $a_l$ referring to the $k'^{th}$ desired symbol and the $l^{th}$ ISI symbol, respectively. $H$ is the raised cosine filter function. $\tau_{err_i}$ is equal to $(\tau_1 + err_i)$ and follows a triangular distribution $\Lambda$ between $-\frac{T_s}{Q}$ and $\frac{T_s}{Q}$.

*B. Impact of imperfect replicas combination on the SNIR*

In order to evaluate the impact of imperfect replicas combination on the performance of MARSALA, we compute the average equivalent SNIR of $y_{sum}$. Therefore, we first derive the power of the desired signal $P(y_{sum,des})$ as shown in the following.

$$P(y_{sum,des}) = \frac{1}{L}\sum_{k'=0}^{L-1} |y_{sum,des}(k'T_s)|^2$$
$$= H^2(\tau_1) + \sum_{i=2}^{N_b} H^2(\tau_{err_i})\left[cos^2(\phi_{err_i}) + sin^2(\phi_{err_i})\right]$$
$$+ 2\sum_{i,j\neq i}^{N_b} H(\tau_{err_i})H(\tau_{err_j})cos(\phi_{err_i})cos(\phi_{err_j})$$
$$+ 2\sum_{i,j\neq i}^{N_b} H(\tau_{err_i})H(\tau_{err_j})sin(\phi_{err_i})sin(\phi_{err_j})$$
$$+ 2H(\tau_1)\sum_{i=2}^{N_b} H(\tau_{err_i})cos(\phi_{err_i}) \quad (15)$$

Given that $\tau_{err_i}$ and $\phi_{err_i}$ are random variables, we compute the average power of the desired signal $E\left[P(y_{sum,des})\right]$. Since the raised cosine filter function can be approximated with a sine cardinal ($sinc$) function on the interval $[-T_s, T_s]$, the mean values of the various terms in (15) are derived as detailed below.

$$E\left[H(\tau_1)\right] = \frac{2Q}{T_s}\int_0^{\frac{T_s}{2Q}} sinc\left(\pi\frac{\tau_1}{T_s}\right) d\tau_1$$
$$= \frac{2Q}{\pi}Si\left(\frac{\pi}{2Q}\right) \quad (16)$$

With $Si$ being the sine integral function.

$$E\left[H^2(\tau_1)\right] = \frac{2Q}{T_s}\int_0^{\frac{T_s}{2Q}} sinc^2\left(\pi\frac{\tau_1}{T_s}\right) d\tau_1$$
$$= \frac{2Q}{\pi}\left(Si\left(\frac{\pi}{Q}\right) - \frac{sin^2\left(\frac{\pi}{2Q}\right)}{\frac{\pi}{2Q}}\right) \quad (17)$$

Assuming that $\tau_1$, $\tau_{err_i}$ and $\tau_{err_j}$ are independant random variables and given that $\tau_{err_i} \sim \Lambda\left(\frac{-T_s}{Q}, \frac{T_s}{Q}\right)$, then we can derive $E[H(\tau_{err_i})]$, $E[H^2(\tau_{err_i})]$ and $E[H(\tau_{err_i})H(\tau_{err_j})]$ as shown in (18), (19) and (20) respectively.

$$E\left[H(\tau_{err_i})\right] =$$

$$\frac{2Q}{T_s}\left[\int_{-\frac{T_s}{Q}}^{0}\left(\frac{Q\tau_{err_i}}{T_s}+1\right)sinc\left(\frac{\pi\tau_{err_i}}{T_s}\right)d\tau_{err_i}\right]$$

$$=\frac{2Q}{\pi}\left(Si\left(\frac{\pi}{Q}\right)+\frac{Q}{\pi}\left(cos\left(\frac{\pi}{Q}\right)-1\right)\right) \quad (18)$$

$$E\left[H^2(\tau_{err_i})\right] =$$

$$\frac{2Q}{T_s}\left[\int_{-\frac{T_s}{Q}}^{0}\left(\frac{Q\tau_{err_i}}{T_s}+1\right)sinc^2\left(\frac{\pi\tau_{err_i}}{T_s}\right)d\tau_{err_i}\right]$$

$$=\frac{2Q}{\pi}\left(Si\left(\frac{2\pi}{Q}\right)-\frac{sin^2\left(\frac{\pi}{Q}\right)}{\frac{\pi}{2Q}}\right)$$

$$+\frac{Q^2}{\pi^2}\left(Ci\left(\frac{2\pi}{Q}\right)-\gamma+log\left(\frac{Q}{2\pi}\right)\right) \quad (19)$$

With $Ci$ being the cosine integral function, $log$ the natural logarithm and $\gamma$ the Euler-Mascheroni constant, $\gamma = 0.577216$.

$$E\left[H(\tau_{err_i})H(\tau_{err_j})\right] =$$
$$\frac{4Q^2}{\pi^2}\left(Si\left(\frac{\pi}{Q}\right)+\frac{Q}{\pi}\left(cos\left(\frac{\pi}{Q}\right)-1\right)\right)^2 \quad (20)$$

The phase difference error $\phi_{err_i}$ can be approximated to a Gaussian variable of mean zero and variance $\sigma^2_{\phi_{err_i}}$ (see Section V-A). Then, the average of $e^{j\phi_{err_i}}$ is equal to $e^{-\frac{\sigma^2_{\phi_{err_i}}}{2}}$, and the variance of $e^{j\phi_{err_i}}$ is $(e^{-\sigma^2_{\phi_{err_i}}}-1)e^{-\sigma^2_{\phi_{err_i}}}$.

To analyze the ISI term after signal combination, we choose to proceed as done in [12] for cooperative MISO systems with time synchronization errors. The raised cosine pulse is approximated to a piecewise linear function with slopes $m_l$. The upper bound of the worst case inter-symbol interference is obtained by considering $a_l H((k'-l)T_s+\tau) = |m_l|\frac{|\tau|}{T_s}$. Thus the ISI term can be written as shown below.

$$y_{sum,isi} = \left(\sum_{l=m-3,l\neq m}^{m+3}|m_l|\right)\frac{|\tau_1|}{T_s}$$
$$+\sum_{i}\underbrace{\left(\sum_{l=m-3,l\neq m}^{m+3}|m_l|\right)}_{\beta}\frac{|\tau_{err_i}|}{T_s}e^{j\phi_{err_i}} \quad (21)$$

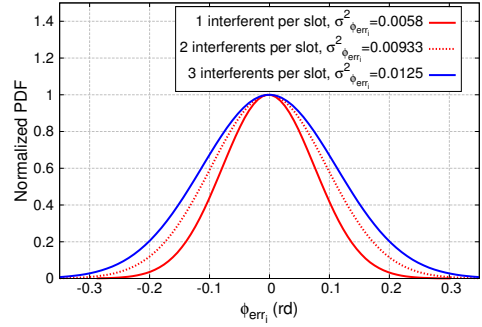

Fig. 2: Normalized PDF of $\phi_{err_i}$ for various numbers of interferents per slot

Since $\tau_1 \sim \mathcal{U}\left(\frac{-T_s}{2Q},\frac{T_s}{2Q}\right)$ and $\tau_{err_i} \sim \Lambda\left(\frac{-T_s}{Q},\frac{T_s}{Q}\right)$, then $|\tau_1| \sim \mathcal{U}\left(0,\frac{T_s}{2Q}\right)$ and $|\tau_{err_i}| \sim \Lambda\left(0,\frac{T_s}{Q}\right)$. The mean and variance values of $y_{sum,isi}$ are given in the following equations respectively.

$$E\left[y_{sum,isi}\right] = \frac{\beta}{4Q}+\sum_{i=2}^{N_b}\frac{\beta}{3Q}e^{-\frac{\sigma^2_{\phi_{err_i}}}{2}} \quad (22)$$

$$var\left[y_{sum,isi}\right] = \frac{\beta^2}{48Q^2}+\sum_{i=2}^{N_b}\frac{\beta^2}{18Q^2}(e^{-\sigma^2_{\phi_{err_i}}}-1)e^{-\sigma^2_{\phi_{err_i}}} \quad (23)$$

The average power of the worst case ISI term is given by

$$E\left[P\left(y_{sum,isi}\right)\right] = E\left[|y_{sum,isi}|^2\right]$$
$$= var\left[y_{sum,isi}\right]+E^2\left[y_{sum,isi}\right] \quad (24)$$

Thus the average equivalent SNIR after imperfect signal combination (SC), due to imperfect synchronisation correction, can be expressed as follows

$$E[SNIR_{eq}] = \frac{E\left[P\left(y_{sum,des}\right)\right]}{E\left[P\left(y_{sum,isi}\right)\right]+(I_1+N_0)+\sum_{i=2}^{N_b}(I_i+N_0)} \quad (25)$$

with $I_1$ and $I_i$ being the power of the total interference term on $TS_{ref}$ and the timeslot containing the $i^{th}$ replica, respectively.

## V. PERFORMANCE EVALUATION

### A. Numerical evaluation of the equivalent SNIR with imperfect replicas combination

Based on the model presented in the previous section, we compute the Probability Density Function (PDF) of $\phi_{err_i}$ through simulations. The numerical results are plotted in Fig.2. Various scenarios are considered with various numbers of interferent packets on each replica. It is observed that the variable $\phi_{err_i}$ follows a Gaussian distribution of mean $\mu = 0$ and variance $\sigma^2_{\phi_{err_i}}$. The worst case estimation error is obtained in the case of three interferents per slot and $\sigma^2_{\phi_{err_i}} = 0.0125$. Then, we evaluate the average degradation of $E[SNIR_{eq}]$ in (25), caused by imperfect correction of timing offsets and phase shifts between replicas. Numerical results are reported in Fig.3. We suppose that the oversampling factor $Q = 4$, $N_b = 2$ replicas and $E_s/N_0 = 7$ dB. To have a reference curve, we plot $SNIR_{eq_{ref}}$, the equivalent $SNIR$ with no error model (i.e. perfect synchronisation correction) obtained with various interference configurations. Then, for the same interference configurations, we plot the average $SNIR_{eq}$ with phase shift and timing offset errors. Both cases with $\sigma^2_{\phi_{err_i}} = 0$ and $\sigma^2_{\phi_{err_i}} = 0.0125$ (worst case scenario) are shown in Fig.3. We can observe that both curves are approximately the same. We conclude that the impact of $\phi_{err_i}$ is not significant on the equivalent $SNIR$ degradation and the degradation is mainly caused by the timing offsets. We can also observe that the degradation of the equivalent SNIR increases from 0.3 dB to 0.5 dB when $SNIR_{eq_{ref}}$ varies from $-1.3$ dB to 2.3 dB. The results obtained with $N_b = 3$ replicas are approximately similar.

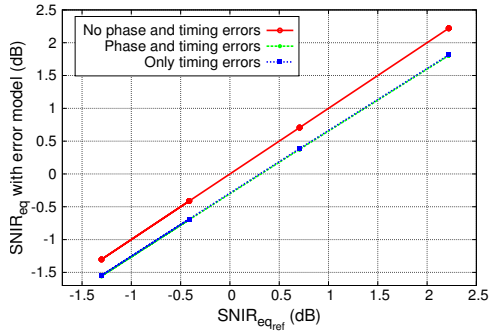

Fig. 3: Equivalent $SNIR$ degradation for incoherent combination of replicas in MARSALA-2, $E_s/N_0 = 7$ dB.

### B. Throughput and PLR simulation results

In the following, we use the numerical results of the analytical model proposed, in order to evaluate the impact of synchronisation errors on the perfomance of MARSALA, in terms of throughput and Packet Loss Ratio (PLR). All the packets are considered equi-powered. The number of terminals transmitting over one frame duration is $\lambda$. Each frame is composed of $N_s = 100$ timeslots. The simulations environment is provided by a satellite communications simulator developped by Thales Alenia Space and CNES. The trafic profile tested is a Constant Bit Rate (CBR) profile. Residual channel estimation errors caused by imperfect interference cancellation have been taken into account. QPSK modulation and DVB-RCS2 turbocode for linear modulation of rate $1/3$ are used in the simulation scenarios. We suppose that the Packet Error Rate (PER) obtained at an SNIR higher than the modcod decoding point is equal to $10^{-5}$, otherwise it is supposed equal to 1. The results obtained for CRDSA have been validated based on the numerical values found in the DVB-RCS2 guidelines. It is worth noting that CRDSA throughput and PLR performance are enhanced when 3GPP turbo code is used instead of DVB-RCS 2 turbo code, however this study will only evaluate the DVB-RCS 2 scheme.

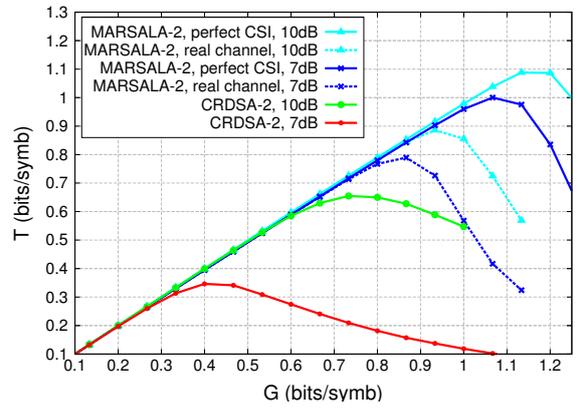

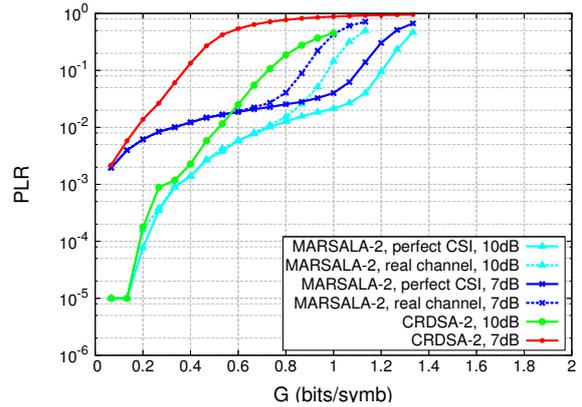

Fig. 4: MARSALA performance with real channel conditions compared to perfect CSI for $E_s/N_0 = 7$ and 10 dB. $N_b = 2$ replicas. QPSK modulation, DVB-RCS2 Turbocode $R = 1/3$. (a) Throughput. (b) PLR.

Once a frame is received, the receiver tries to decode a packet using CRDSA. If the decoding is not successful, the receiver applies MARSALA to localise and combine the replicas of this packet. This procedure is repeated until no more packets can be decoded successfully. In order to compare several modcods, the normalized load (G) is expressed in bits per symbol and computed as shown below

$$G = R * Log_2(M) * \frac{\lambda}{N_s} \quad (26)$$

with R being the code rate and M the modulation order. The normalized throughput (T) is given by

$$T = G(1 - PLR(G)) \quad (27)$$

where PLR(G) is the probability that a packet is not decoded for a given G and a given SNIR. We denote by MARSALA-2 and CRDSA-2, the MARSALA and CRDSA systems where each terminal transmits 2 replicas of a given packet. The same notation is taken for MARSALA-3 and CRDSA-3.

Following the numerical results of the theoretical study in Section IV-B, Fig.4 and Fig.5 trace the throughput and the PLR obtained with MARSALA-2 and MARSALA-3 in case of perfect knowledge of Channel State Information (CSI) as well as real channel conditions. We can observe that, for MARSALA-2 in real channel conditions, the throughput is degraded compared to the perfect CSI case, by around $18\%$ to $20\%$. However, for MARSALA-3, the degradation is around $10\%$ at $\frac{E_s}{N_0} = 3$ and 7 dB, and $6\%$ at $\frac{E_s}{N_0} = 10$ dB. The degradation is less significant for MARSALA-3, because the equivalent SNIR obtained with the combination of 3 replicas is higher and permits more successful decoding than the case of 2 replicas, even in presence of synchronisation errors. As for PLR curves, they show that the best performance of MARSALA in terms of PLR is obtained with $N_b = 3$ replicas and $\frac{E_s}{N_0} = 10$ dB. In this case, the PLR in real channel conditions is only slightly lower then the perfect CSI case (the PLR is the same until $G > 1$ bits/symbol.

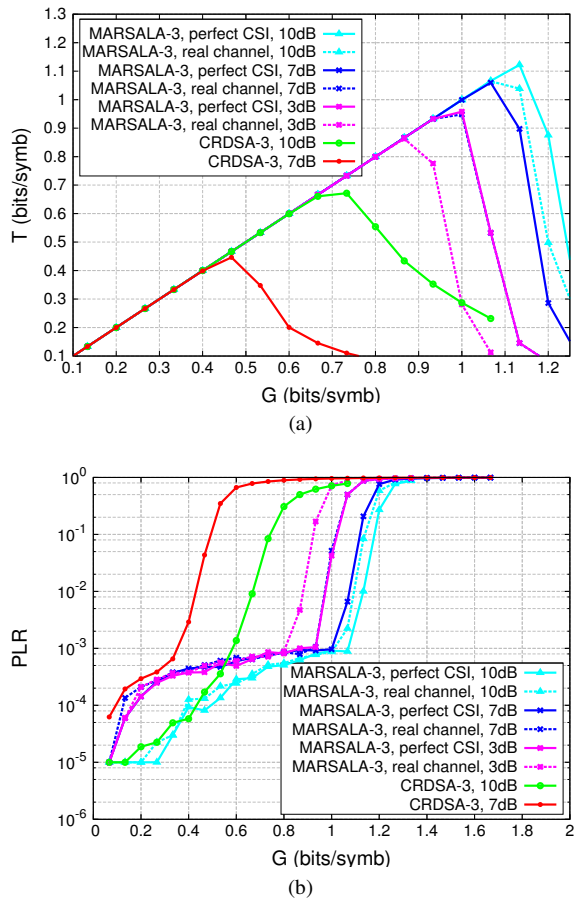

Fig. 5: MARSALA performance with real channel conditions compared to perfect CSI for $E_s/N_0 = 3$, 7 and 10 dB. $N_b = 3$ replicas. QPSK modulation, DVB-RCS2 Turbocode $R = 1/3$. (a) Throughput. (b) PLR.

## VI. CONCLUSION AND FUTURE WORK

In this paper, we have designed a method to estimate and compensate the synchronisation errors between the replicas of a given packet, prior to signal combination in MARSALA. Based on the proposed estimation method, we have presented a detailed analytical model to evaluate the impact of signal misalignment on the combination performance in MARSALA. We have shown that the correction of the timing offsets differences and phase shifts between the replicas of a given packet prior to their combination is a critical task in MARSALA, and it can degrade the performance, if not done accurately. In order to further evaluate MARSALA combined with CRDSA, ongoing work is testing MARSALA with 3GPP turbocode and in the case of unbalanced packets power. Finally, an interesting future work would be to combine MARSALA with Asynchronous Contention Resolution Diversity Aloha (ACRDA) [13], given the overall advantages of the asynchronous access mode.

REFERENCES

[1] L. G. Roberts, "ALOHA Packet System with and without Slots and Capture," *ACM, SIGCOMM Computer Communication Review*, vol. 5, no. Feb., pp. 28–42, 1975.

[2] G. Choudhury and S. Rappaport, "Diversity ALOHA–A Random Access Scheme for Satellite Communications," *IEEE Transactions on Communications*, vol. 31, no. 3, pp. 450–457, Mar 1983.

[3] E. Casini, R. D. Gaudenzi, and O. D. R. Herrero, "Contention Resolution Diversity Slotted ALOHA (CRDSA) : An Enhanced Random Access Scheme for Satellite Access Packet Networks," *IEEE Transactions on Wireless Communications*, 2007.

[4] *DVB Document ETSI A155-1, Digital Video Broadcasting (DVB); Second Generation DVB Interactive Satellite System (RCS2); Part2*, ETSI Std., Rev. 1.1.1, 2011.

[5] *Digital Video Broadcasting (DVB); Second Generation DVB Interactive Satellite System (DVB-RCS2); Guidelines for Implementation and Use of LLS: EN 301 545-2*, ETSI Std., 2012.

[6] O. del Rio Herrero, "A high-performance MAC protocol for Consumer Broadband Satellite Systems," *IET Conference Proceedings*, pp. 512–512(1).

[7] G. Liva, "Graph-Based Analysis and Optimization of Contention Resolution Diversity Slotted ALOHA," *IEEE Transactions on Communications*, vol. 59, no. 2, pp. 477–487, February 2011.

[8] H.-C. Bui, J. Lacan, and M.-L. Boucheret, "An Enhanced Multiple Random Access Scheme for Satellite Communications," in *Wireless Telecommunications Symposium (WTS), 2012*, April 2012, pp. 1–6.

[9] H.-C. Bui, K. Zidane, J. Lacan, and M.-L. Boucheret, "A Multi-Replica Decoding Technique for Contention Resolution Diversity Slotted Aloha," in *Vehicular Technology Conference (VTC Fall), 2015 IEEE 82st*, September 2015, pp. 1–5.

[10] K. Zidane, J. Lacan, M.-L. Boucheret, and C. Poulliat, "Improved Channel Estimation for Interference Cancellation in Random Access Methods for Satellite Communications," in *Advanced Satellite Multimedia Systems Conference and the 13th Signal Processing for Space Communications Workshop (ASMS/SPSC), 2014 7th*, Sept 2014, pp. 73–77.

[11] K. Zidane, J. Lacan, M.-L. Boucheret, C. Poulliat, M. Gineste, D. Roques, C. Bes, and A. Deramecourt, "Effect of Residual Channel Estimation Errors in Random Access Methods for Satellite Communications," in *Vehicular Technology Conference (VTC Spring), 2015 IEEE 81st*, May 2015, pp. 1–5.

[12] S. Jagannathan, H. Aghajan, and A. Goldsmith, "The Effect of Time Synchronization Errors on the Performance of Cooperative MISO Systems," in *IEEE Global Telecommunications Conference Workshops (Globecom), 2004.*, Nov 2004, pp. 102–107.

[13] R. D. Gaudenzi, O. del Ro Herrero, G. Acar, and E. G. Barrabs, "Asynchronous Contention Resolution Diversity ALOHA: Making CRDSA Truly Asynchronous," *IEEE Transactions on Wireless Communications*, vol. 13, no. 11, pp. 6193–6206, Nov 2014.